\documentclass[aip,pop,groupedaddress,reprint]{revtex4-1}
\usepackage{graphicx}
\usepackage{ulem}
\usepackage{amssymb}
\begin{document}

\flushbottom

\title{Transport across nanogaps using semiclassically consistent boundary conditions}
 
\vskip 0.3 in

\author{Debabrata Biswas}
\author{Pradeep Baraila}
\author{Raghwendra Kumar}

\affiliation{Theoretical Physics Division,
Bhabha Atomic Research Centre,
Mumbai 400 085, INDIA}

\date{\today}
\vskip 0.2 in

\begin{abstract}
Charge particle transport across nanogaps is studied theoretically within the Schrodinger-Poisson 
mean field framework and the existence of limiting current investigated.
It is shown that the choice of a first order WKB wavefunction as the transmitted wave 
leads  to self consistent boundary conditions and gives results that are significantly
different in the non-classical regime from those obtained using a plane transmitted
wave. At zero injection energies, the quantum limiting current density ($J_c$) is 
found to obey the local scaling law $J_c \sim V_g^\alpha/D^{5-2\alpha}$ with the 
gap separation $D$ and voltage $V_g$. The exponent $\alpha > 1.1$ with  
$\alpha \rightarrow 3/2$ in the 
classical regime of small de Broglie
wavelengths. These results are consistent with recent experiments 
using nanogaps most of which are found to be in a parameter regime where 
classical space charge limited scaling holds away from the
emission dominated regime. 
\end{abstract}

\vskip 0.25 in

\maketitle

\newcommand{\be}{\begin{equation}}
\newcommand{\ee}{\end{equation}}
\newcommand{\bea}{\begin{eqnarray}}
\newcommand{\eea}{\end{eqnarray}}
\newcommand{\Tbar}{{\bar{T}}}
\newcommand{\En}{{\cal E}}
\newcommand{\Lop}{{\cal L}}
\newcommand{\DB}[1]{\marginpar{\footnotesize DB: #1}}
\newcommand{\q}{\vec{q}}
\newcommand{\kt}{\tilde{k}}
\newcommand{\Lopn}{\tilde{\Lop}}
\newcommand{\noi}{\noindent}
\newcommand{\ovn}{\bar{n}}
\newcommand{\ovx}{\bar{x}}
\newcommand{\ovE}{\bar{E}}
\newcommand{\ovV}{\bar{V}}
\newcommand{\ovU}{\bar{U}}
\newcommand{\ovJ}{\bar{J}}
\newcommand{\calE}{{\cal E}}

The current-voltage characteristics across nanogaps is a subject of much interest
across a wide variety of fields. An important quantity is the maximum 
current that can be transmitted across the gap. In classical physics, the limit exists 
due to the mutual repulsion between the charged particles and the 
current is said to be space charge limited. In a mean
field picture, the mutual repulsion gives rise to a potential barrier.
As the current reaches  the limiting value, the barrier height increases and 
the transmitted particles barely cross the barrier. Beyond this value,
the barrier height oscillates, some electrons get reflected and steady state transport 
is no longer possible. 
The Child-Langmuir \cite{child,langmuir} law and its many generalizations \cite{lau2d,lugin,pop3} 
serve well to predict the
maximum transmitted current in macro devices where the available phase space volume
is large and quantum effects can be safely ignored. 

In nanostructures however,
the small physical size together with low applied voltages or injection energies
pushes the average per electron phase space volume towards the plank cell limit.
Thus, in addition to space charge, quantum effects must be considered
in exploring the existence of such a limit in nanoelectronics, vacuum microelectronics
or devices such as the scanning tunneling microscopes. In particular, the possibility
of tunneling leads to the question: {\it is there a limit on the maximum transmitted current density
in the quantum mechanical case} ?
Theoretical studies  \cite{lau,ang2003} suggest that quantum 
tunneling pushes up this limit and by orders of 
magnitude in the low injection energy regime where quantum effects 
should be considerably more important. In fact, within the framework
considered in Ref. \onlinecite{lau}, the limit 
can indeed be very large in the {\it very low energy} regime as we shall 
show here. This singular behavior,
signalling a sharp departure from the classical prediction, warrants a fresh look
at the basic assumptions involved, especially since, many of them are retained in
more sophisticated theories \cite{ang2003,nano,ang_pop} which take into account 
the fermionic nature of
electrons by incorporating the exchange-correlation potential.

In its simplest form, the mean field quantum (Hartree) formalism involves solving the 
coupled Schrodinger and Poisson
equations

\bea
&-& \frac{\hbar^2}{2m} \frac{d^2\psi}{dx^2} - eV(x) \psi = E\psi \label{eq:Sch0}\\
&~& \frac{d^2 V}{dx^2} = \frac{e|\psi|^2}{\epsilon_0}
\eea

\noi
in the region $[0,D]$ 
where $D$ is the size of the gap, $e$ is the magnitude of the electronic charge, $E$
is the energy of the electron and $V(0) = 0$ while $V(D) = V_g$. Such a model
is expected to hold when the electron density in the gap is sufficiently 
high that their mutual interaction must be accounted for, but is low enough 
to neglect the effects of the exclusion principle. It is assumed here that $V$
is time-independent and there is a  steady current flowing across the gap 
with a current density $J = e i\hbar ( \psi^* \psi^{'} - \psi \psi^{*'}) / {2m}$.
The boundary conditions or initial values for $\psi$ are to be determined 
taking into account this fact.

Following Ref. \onlinecite{lau}, we write the wavefunction $\psi$ as 

\be
\psi(x) = (n_s \ovE)^{1/2}~~r(\ovx)e^{i\theta(\ovx)}
\ee

\noi
where $r(\ovx)$ and $\theta(\ovx)$ are real, $\ovE = E/eV_s$ is a dimensionless energy
and $n_s$ is a characteristic density.
In terms of other dimensionless quantities $\bar{x} = x/D$,
$\bar{V} = eV/E$, $\ovJ = J/J_s$, $\bar{n} = n/n_s = |\psi|^2/n_s$
and $\phi_g = eV_g/E$, where $V_s = \hbar^2/(2meD^2)$, 
$J_s = \epsilon_0 \hbar^3/(4m^2eD^5)$, $n_s = \epsilon_0 \hbar^2/(2me^2D^4)$, the 
coupled Schrodinger and Poisson equations can be expressed as

\bea
&~& \frac{d^2 r}{d\ovx^2} + \ovE [(1 + \ovV) - \frac{(\lambda/4)^2}{r^4} ]r = 0 \label{eq:Sch}\\
&~& \frac{d^2 \ovV}{dx^2} = r^2 \label{eq:Poi} \\
&~& \frac{d \theta}{d\ovx} = \frac{\lambda\ovE^{1/2}}{4r^2(\ovx)} 
\eea

\noi
where $\lambda = 2\ovJ/\ovE^{3/2}$ is a dimensionless perveance. Note that
once $r(\ovx)$ is known, $\theta(\ovx)$ can be determined independently 
with an arbitrary phase ($\theta(1) = 0$) at the boundary $x = D$.

Assuming now that at $x \simeq D$, $\psi(x) = C e^{ip(D)x/\hbar}$, it follows
that $r(1) = (\lambda/4)^{1/2}/(1 + \phi_g )^{1/4}$ while $r'(1) = 0$. Here
$p(x) = \sqrt{2m(E + eV(x))}$ is the classical momentum.
Note that the plane wave assumption above presupposes a constant potential immediately
beyond the domain of interest \cite{lau} irrespective of the nature of the interface
at $x = D$, the length scales involved and the self consistent 
nature of the problem. 

At a given scaled injection energy $\ovE$, the dimensionless
perveance $\lambda$ is increased till no solution exists. This gives the critical 
current density $J_c = J_s \lambda_c \ovE^{3/2} /2$. In Ref.~\onlinecite{lau}, it has
been observed that for small values of $\ovE$, $\lambda_c$
exceeds the classical values by orders of magnitude. 

The behaviour of the system at low injection energies ($\ovE$) can be analysed
by neglecting the second term altogether in Eq.~(\ref{eq:Sch}) at finite values
of $\lambda$.  It is then easy to see that  at very low energies, a solution
exists as $\lambda$ is increased atleast so long as the second term in Eq.~(\ref{eq:Sch})
can be neglected. This behaviour is however peculiar to the initial conditions
specified above. We show later that the same equations albeit 
with a different initial condition for $r'(1)$  imposes a limit on $\lambda$ 
beyond which no solution exists even at low injection energies.

Apart from the $\ovE \rightarrow 0$ behaviour,  
the semiclassical consistency of the transmitted plane wave assumption is also
worth investigating. 
If the potential $\ovV$ for a particular value of $\ovE$ and $\lambda$ is
well behaved in the region $[0,D]$, it is reasonable to expect that
away from classical turning points of the potential $\ovV$, the transmitted 
wavefunction should have the standard first order WKB form:

\be
\psi_{sc}(x) = \frac{C}{\sqrt{p(x)}} e^{\frac{i}{\hbar}\int^x p(x') dx' + i \varphi} \label{eq:semi}
\ee

\noi
as $x$ approaches $D$. Here $p(x)$ is the classical
momentum while $C$ is assumed to be real with all the
phase information dumped in $\varphi$. 
Assuming this form to be true at $x = x_0 < D$,
it is possible to compare $r(\ovx)$ and $\theta(\ovx)$ with the 
semiclassical amplitude and phase by integrating backwards from
$\ovx_0 = x_0/D$. Thus, starting with $r(\ovx_0) =  r_{sc}(\ovx_0) = C/\sqrt{p(\ovx_0)}$ and
$\theta(\ovx_0) = \theta_{sc}(\ovx_0) = \int^{x_0} p(x') dx'/\hbar +  \varphi$, 
one can compare $r(\ovx)$ and
$\theta(\ovx)$ for values of $x < x_0$ with the semiclassical (WKB) 
predictions: $r_{sc}(\ovx) =  C/\sqrt{p(\ovx)}$ and  $\theta_{sc}(\ovx) =  \int^x p(x') dx'/\hbar +  \varphi
= \theta(\ovx_0) - \int_x^{x_0} p(x') dx'/\hbar$. A comparison of the amplitudes
for $\ovE = 1$ and $1000$
is shown in Fig.~1 for $\lambda = 15$ and $\phi_g = 0.5$. While $\theta(\ovx)$ and
$\theta_{sc}(\ovx)$ agree reasonably at $\ovE = 1$ near $\ovx_0 = 0.9$ (not shown here), 
the amplitude $r(\ovx)$ has very little agreement 
with $r_{sc}(\ovx)$ even slightly away from $\ovx = \ovx_0$ (see Fig.~1).
At $\ovE = 1000$ however, the amplitude oscillates about the
the semiclassical prediction while $\theta(\ovx)$ and
$\theta_{sc}(\ovx)$ are practically indistinguishable in the entire region.

\begin{figure}[tbh]
\begin{center}
\includegraphics[width=5cm,angle=270]{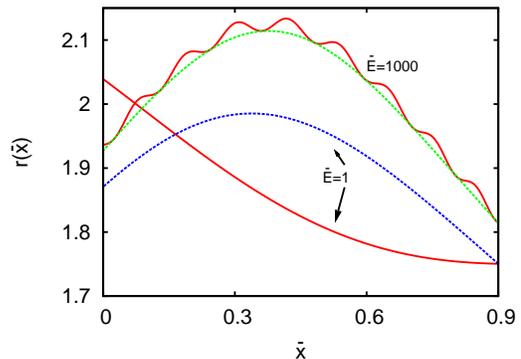}
\end{center}
\caption{(Color online) A comparison of the amplitude $r(\ovx)$ obtained using the plane wave
boundary condition with the semiclassical prediction $r_{sc}(\ovx)$ (dashed curves). 
The agreement is poor at $\ovE = 1$ but improves for $\ovE=1000$. Here $\lambda = 15$ 
and $\phi_g = 0.5$ while $\ovx_0=0.9$. }
\label{fig:1}
\end{figure}

These results are indeed not very surprising since the plane wave
approximation is only the first term in a semiclassical expansion
of $\psi(x) = e^{\frac{i}{\hbar}S(x)}$: 
$S(x)  =  S_0(x) + \frac{\hbar}{i}S_1(x) + (\frac{\hbar}{i})^2 S_2(x) + \ldots$
where $S_0(x) = \int^x p(x') dx'$. By dropping $S_1$ and subsequent terms in 
a plane wave expansion, it is only to be expected that the amplitude 
is not reproduced accurately even for large $\ovE$ while the phase information
is well approximated as $\ovE$ increases.

For a well behaved potential, a better approximation for the transmitted 
wave near $x = D$ 
should be the first order WKB wavefunction in  Eq.~(\ref{eq:semi}), while
for $x \simeq 0$, it should be a superposition of a right and a left moving
wave of the WKB form: 
$\psi_{sc}(x \simeq 0) = (A/\sqrt{p(x)}) e^{\frac{i}{\hbar}\int^x p(x') dx'} + 
(B/\sqrt{p(x)})  e^{-\frac{i}{\hbar}\int^x p(x') dx'} \label{eq:semi1}$. 

\begin{figure}[tbh]
\begin{center}
\includegraphics[width=4.5cm,angle=270]{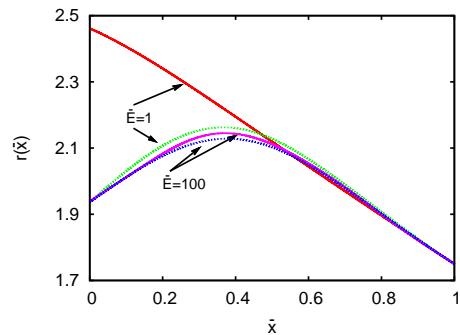}
\end{center}
\caption{(Color online) A comparison of the amplitude $r(\ovx)$ obtained using the new initial condition 
with the WKB prediction $r_{sc}(\ovx)$ for $\lambda = 15$ and $\phi_g = 0.5$. 
The agreement is better even at $\ovE = 1$. At
$\ovE=100$, there is practically no reflection. The agreement is 
therefore good even near $\ovx=0$.
The solid lines are the exact results; the dashed/dotted lines are the WKB predictions.
}
\label{fig:2}
\end{figure}

Assuming the semiclassical form (Eq.~(\ref{eq:semi})) to be valid at $x = D$, the
initial conditions for $r(\ovx)$ are $r(1) = (\lambda/4)^{1/2}/(1 + \phi_g )^{1/4}$ 
and $r'(1)  = - \ovV'(1) \lambda^{1/2}/[8(1 + \phi_g )^{5/4}]$. 
Note that $r'(1)$ depends on $\ovV'(1)$. This does not preclude $r'(1) = 0$; rather
it forces the potential to assume a constant value smoothly at $x \simeq D$ in order
to be semiclassically consistent \cite{semi_boundary,singer}.
In practice, Eqns. (\ref{eq:Sch}) and (\ref{eq:Poi}) are solved as an initial
value problem starting at $\ovx = 1$ and  integrating backwards till $\ovx = 0$
by choosing a value for $\ovV'(1)$ such that $\ovV(0) = 0$ \cite{2_solutions}. For the
parameter values studied by us however, $\ovV'(1) \neq 0$ for any allowed
solution set.   
Not surprisingly, a comparison of $r(\ovx)$ and $r_{sc}(\ovx)$ shows a much better
agreement now as shown in Fig.~2.

Imposition of the semiclassical boundary condition also removes the
singular behaviour at low scaled energies ($\ovE \rightarrow 0$). 
Neglecting the second term in Eq.~(\ref{eq:Sch})), and using the new initial
conditions, the amplitude equation gives 
$r(\ovx) = ax + b$ where 
$a + b  =  (\lambda/4)^{1/2}/(1 + \phi_g )^{1/4} \label{eq:c1}$ and
$a  =   - \ovV'(1) \lambda^{1/2}/[8(1 + \phi_g )^{5/4}]$.
Inserting $r(\ovx)$ in the Poisson equation (Eq.~\ref{eq:Poi}) and using the boundary 
condition  $\ovV(0)=0$ gives $\ovV(\ovx) = a^2\ovx^4/12 + ab\ovx^3/3 + b^2\ovx^2/2 + c\ovx$.
Finally on using $\ovV(1)=\phi_g$ and demanding that 
$r'(1) = - (a^3/3 + ab + b^2 + c) \lambda^{1/2}/[8(1 + \phi_g )^{5/4}] $, a real
solution is found to exist only when $(48\calE^6)^2 - (96\calE^6 + 12\calE^2\phi_g)\lambda - \lambda^2/2 > 0$
where $\calE = (1 + \phi_g)^{1/4}$. This imposes an upper limit on the perveance $\lambda$ at low
injection energies which matches with the critical perveance $\lambda_c$ at small $\ovE$ obtained
by solving the full set of equations (Eqns.~(\ref{eq:Sch}) and (\ref{eq:Poi})) for 
different values of $\lambda$. For instance, the inequality above predicts that
a real solution exists at $\phi_g= 0.5$ for $\lambda < 38.33$ at very small values of
$\ovE$. This is a good approximation to the actual value at small $\ovE$ as
can be seen in Fig.~3 where $\lambda_c$ is plotted for different scaled injection energies
for  $\phi_g$ =  0.5, 0 and -0.5. Note that $\lambda_c$ does not
increase by orders of magnitude even at low injection energies. Rather the quantum regime
manifests itself differently at various applied potentials. While $\lambda_c$ is about 
double the classical value at $\phi_g = 0.5$, it is less than the 
classical value at $\phi_g = -0.5$ where reflection (rather than tunneling) dominates. 
In general, at most injection energies, the maximum current evaluated quantum
mechanically falls short of the classical prediction.

\begin{figure}[tbh]
\begin{center}
\includegraphics[width=4.5cm,angle=270]{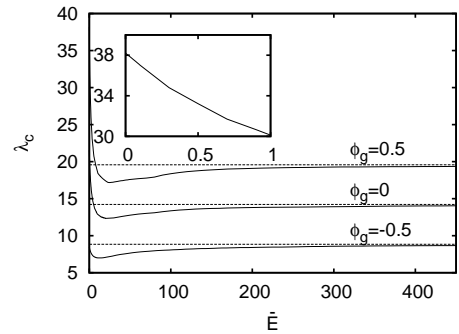}
\end{center}
\caption{The critical perveance $\lambda_c$ is plotted against $\ovE$. The dashed lines are
the classical prediction while the solid lines are obtained by numerically
solving Eqns.~(\ref{eq:Sch}) and (\ref{eq:Poi}). Inset shows the value of $\lambda_c$ at
small $\ovE$ for $\phi_g = 0.5$.}
\label{fig:3}
\end{figure}

\begin{figure}[htb]
\begin{center}
\includegraphics[width=4.5cm,angle=270]{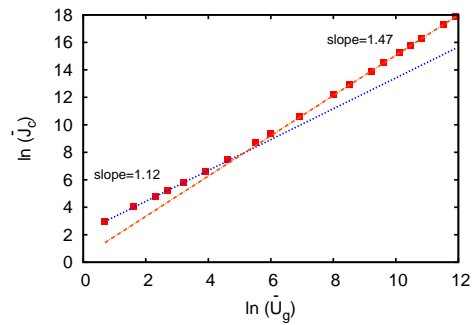}
\end{center}
\caption{(Color online) The scaled critical current $\ovJ_c$ follows a local power law relationship
with the scaled applied voltage: $\ovJ_c \sim \ovU_g^\alpha$. Boxes mark the 
numerically obtained values while the short and long  dashed lines are the
fits at small and large $\ovU_g$ respectively. At small $\ovU_g$, 
$\alpha \simeq 1.12$
while for large $\ovU_g$, $\alpha \rightarrow 1.5$.}

\label{fig:J_vs_U}
\end{figure}

When the injection energy is exactly zero, ($E=0$ in Eq.~(\ref{eq:Sch0})), the semiclassical 
formalism in the space charge limited regime is similar but the equations and 
initial conditions are slightly different \cite{archival}.  It is interesting to investigate whether the
scaling relationship with applied voltage
is different from the classical $V_g^{3/2}$ law of Child-Langmuir with the new
boundary condition. The variation of the scaled critical current 
$\ovJ_c= J_c/J_s$ with $\ovU_g =  V_g/V_s$  is 
shown in Fig.~\ref{fig:J_vs_U}. Clearly, $\ovJ_c \sim \ovU_g^\alpha$ locally
with the exponent in the range $1.1 \le \alpha \leq 1.5$ and converging to 1.5 as 
$\ovU_g$ is increased. Note that $\ovU_g = V_g/V_s \sim V_g D^2$ while 
$\ovJ_c = J_c/J_s \sim J_c D^5$. Thus $J_c D^5 \sim V_g^\alpha D^{2\alpha}$ or 

\be
J_c \sim \frac{V_g^\alpha}{D^{5 - 2\alpha}}
\ee
 
\noi
Thus, as $V_g$ and $D$ are increased, $J_c$ obeys classical scaling. Note that
the $\alpha=1/2$ behaviour reported in  Ref. \onlinecite{ang2003} is not
seen in our computations with the first order WKB boundary conditions.
It is in fact possible to show \cite{archival} that the plane wave boundary condition
predicts $\alpha = 1/2$. To see this, note that $\ovJ \sim \ovU_g^{1/2} \ovn(1)$ 
where $\ovn(1) = q^2(1)$ and $\psi(x) = q(\ovx) \sqrt{n_s} e^{i\theta(\ovx)}$.
Since 

\be
\frac{d\ovn}{d\ovU}|_{\ovU=\ovU_g} = \frac{d\ovn}{d\ovx}|_{\ovx=1} \frac{d\ovx}{d\ovU}|_{\ovU=\ovU_g}
\ee

\noi
and $q'(1) = 0$ for a plane wave
boundary condition, $d\ovn/d\ovU|_{\ovU=\ovU_g} = 0$. Thus, $\ovn(1)$ is 
independent of $\ovU_g$. It follows that  
$J \sim V_g^{1/2}/D^4$ only if the transmitted wavefunction is assumed 
to be a plane wave.

Note that  the smallest value \cite{higher_order} of the scaled potential 
considered is $\ovU_g = 2$.
At $D=10$nm, this translates to an applied potential of $0.75$mV while at
$D=70$nm, it is $0.015$mV. On the higher 
side where classical scaling holds, the maximum scaled potential
considered is $\ovU_g = 150000$. At $D=10$nm, the maximum $V_g = 57$V, while at
$D=70$nm, the maximum applied voltage considered is $V_g = 1.16$V. Beyond these
voltages, classical scaling should be applicable. These results are consistent with 
recent experiments on aluminium \cite{iitk09} and graphene \cite{graphene} nanogaps.
In case of aluminium for instance (see fig.~3 of Ref. \onlinecite{iitk09}), 
the $70$nm gap shows good agreement
with the Fowler-Nordheim emission law for $V_g \lessapprox 6$V. Above this
(see Fig.~4 of Ref. \onlinecite{iitk09}), the current
appears to be space charge limited with the exponent $\alpha \simeq 1.5$  which
is consistent with the first order WKB results presented here.

In case of graphene \cite{graphene}, the space charge limited regime occurs at 
smaller voltages possibly on account of Klein tunneling\cite{klein}. 
For a few hundred nanometer gap, the exponent
$\alpha$ averaged over several gate voltages, takes the classical space charge 
limited value of $1.5$ for $V_g \gtrapprox 3$V. This is again consistent
with the results presented here.

In conclusion, we have shown that a first order WKB approximation for the
transmitted wave gives semiclassically consistent results which are
significantly different from those obtained using a plane transmitted wave.
The maximum transmitted current  
is found to be smaller than the classical current for most
injection energies and  may exceed only in the quantum limit of small $\ovE$.
Finally, at zero injection energy, the limiting current density follows a 
local scaling law $J_c \sim V_g^\alpha/D^{5 - 2\alpha}$ 
with $\alpha \rightarrow 3/2$ for large $V_g$ and $D$. Importantly, our
result using first order WKB approximation \cite{higher_order} underscores
the possible parameter regime where nonclassical space charge limited 
behaviour may be observed experimentally.

\newcommand{\PR}[1]{{Phys.\ Rep.}\/ {\bf #1}}
\newcommand{\PRL}[1]{{Phys.\ Rev.\ Lett.}\/ {\bf #1}}
\newcommand{\PP}[1]{{Phys.\ Plasmas\ }\/ {\bf #1}}
\newcommand{\JAP}[1]{{J.\ Appl.\ Phys.}\/ {\bf #1}}



\begin{thebibliography}{99}
\bibitem{semi_boundary} Semiclassical boundary conditions have been used before; see
\cite{singer}
\bibitem{2_solutions} There are two sets of solutions which
merge as the critical current is approached. We choose the one with the lower potential
energy.
\bibitem{archival} The details will be presented in a separate communication.
\bibitem{higher_order} At very small spacings or applied voltages, higher order WKB wavefunctions
may be required.
\end{thebibliography}
\end{document}